\documentclass{emulateapj}

\def\ts     {\thinspace}
\def\kms    {\ts km\ts s$^{-1}$}
\def\etal   {{\rm et\ts al.}}
\def\msol   {$M_{\odot}$}
\def\lsol   {$L_{\odot}$}

\def\aco    {{\rm CO}($J$=1$\to$0)}
\def\bco    {{\rm CO}($J$=2$\to$1)}

\def\eco    {{\rm CO}($J$=5$\to$4)}

\def\bri    {BRI\,1335--0417}

\slugcomment{draft version \today, accepted for publication in the Astrophysical Journal Letters}

\shorttitle{High--Resolution CO(2--1) Imaging of BRI\,1335--0417}
\shortauthors{Riechers et al.}

\begin{document}

\title{
  Formation of a Quasar Host Galaxy through a Wet Merger \\ 1.4 Billion
  Years after the Big Bang}

\author{Dominik A. Riechers\altaffilmark{1,2,5}, Fabian Walter\altaffilmark{1}, 
Christopher L. Carilli\altaffilmark{3}, Frank Bertoldi\altaffilmark{4}, 
and Emmanuel Momjian\altaffilmark{3}}

\altaffiltext{1}{Max-Planck-Institut f\"ur Astronomie, K\"onigstuhl 17, 
Heidelberg, D-69117, Germany}

\altaffiltext{2}{Astronomy Department, California Institute of
  Technology, MC 105-24, 1200 East California Boulevard, Pasadena, CA
  91125; dr@caltech.edu}

\altaffiltext{3}{National Radio Astronomy Observatory, PO Box O, Socorro, NM 87801}

\altaffiltext{4}{Argelander-Institut f\"ur Astronomie, Universit\"at Bonn, Auf dem
  H\"ugel 71, Bonn, D-53121, Germany}

\altaffiltext{5}{Hubble Fellow}


\begin{abstract}
  We present high-resolution Very Large Array imaging of the molecular
  gas in the host galaxy of the high redshift quasar \bri\
  ($z=4.41$). Our \bco\ observations have a linear resolution of
  0.15$''$ (1.0\,kpc) and resolve the molecular gas emission both
  spatially and in velocity. The molecular gas in \bri\ is extended on
  scales of 5\,kpc, and shows a complex structure.  
  At least three distinct components encompassing about two thirds of
  the total molecular mass of 9.2$\times$10$^{10}$\,\msol\ are
  identified in velocity space, which are embedded in a structure that
  harbors about one third of the total molecular mass in the
  system. The brightest \bco\ line emission region has a peak
  brightness temperature of 61$\pm$9\,K within 1\,kpc diameter, which
  is comparable to the kinetic gas temperature as predicted from the
  CO line excitation.  This is also comparable to the gas temperatures
  found in the central regions of nearby ultra-luminous infrared
  galaxies, which are however much more compact than 1\,kpc. The
  spatial and velocity structure of the molecular reservoir in \bri\
  is inconsistent with a simple gravitationally bound disk, but
  resembles a merging system.  Our observations are consistent with a
  major, gas-rich (`wet') merger that both feeds an accreting
  supermassive black hole (causing the bright quasar activity), and
  fuels a massive starburst that builds up the stellar bulge in this
  galaxy.  Our study of this $z$$>$4 quasar host galaxy may thus be
  the most direct observational evidence that `wet' mergers at high
  redshift are related to AGN activity.
\end{abstract}



\section{Introduction}

Great progress has been made in recent years both observationally and
theoretically to further our understanding of galaxy formation and
evolution from the early epochs of galaxy formation to the present-day
universe.  One basic prediction of cosmological simulations is that
during the early epoch of hierarchical galaxy formation, some of the
most massive galaxies are already formed in major merger events (e.g.,
Springel \etal\ \citeyear{spr05}). These mergers are believed to
commonly trigger both AGN and starburst activity in such early
systems, which is regulated via AGN feedback (e.g., Hopkins \etal\
\citeyear{hop05}). Such feedback may be responsible for the
present-day ``$M_{\rm BH}$--$\sigma_v$'' relation between black hole
mass and bulge velocity dispersion (Ferrarese \& Merritt
\citeyear{fer00}; Gebhardt \etal\ \citeyear{geb00}). 

Studies of molecular gas (the requisite material to fuel star
formation) in young galaxies are an essential ingredient to understand
their physical properties in more detail.
Molecular gas (typically CO) has been detected in $\sim$50 galaxies at
$z$$>$1 to date, revealing large molecular reservoirs of
$>$10$^{10}$\,\msol\ in most cases (see Solomon \& Vanden Bout
\citeyear{sv05} for a review). However, these studies rely almost
exclusively on the integrated properties of the line emission, as the
molecular reservoirs in these distant galaxies are difficult to
resolve. To date, only the $z$=4.69 and $z$=6.42 quasars
BR\,1202--0725 and SDSS\,J1148+5251, two of the most distant gas-rich
galaxies could be resolved in molecular gas emission without the aid
of gravitational magnification (Omont \etal\ \citeyear{omo96}; Carilli
\etal\ \citeyear{car02}; Walter \etal\ \citeyear{wal04}).

In this letter, we report on high angular resolution (0.15$''$;
1.0\,kpc) Very Large Array (VLA)\footnote{The Very Large Array is a
facility of the National Radio Astronomy Observatory, operated by
Associated Universities, Inc., under a cooperative agreement with the
National Science Foundation.}  observations of molecular gas
in the host galaxy of BRI\,1335--0417, a dust-rich, optically
identified quasar at a redshift of 4.41, corresponding to only 1.4 Gyr
after the Big Bang.
Optical imaging with the Hubble Space Telescope reveals a single point
source without any evidence for
gravitational lensing (Storrie-Lombardi \etal\ \citeyear{sl96}).  We
use a concordance, flat $\Lambda$CDM cosmology throughout, with
$H_0$=71\,\kms\,Mpc$^{-1}$, $\Omega_{\rm M}$=0.27, and
$\Omega_{\Lambda}$=0.73 (Spergel \etal\ 
\citeyear{spe07}).

\section{Observations}

\begin{figure}
\epsscale{1.15}
\vspace{-5mm}

\plotone{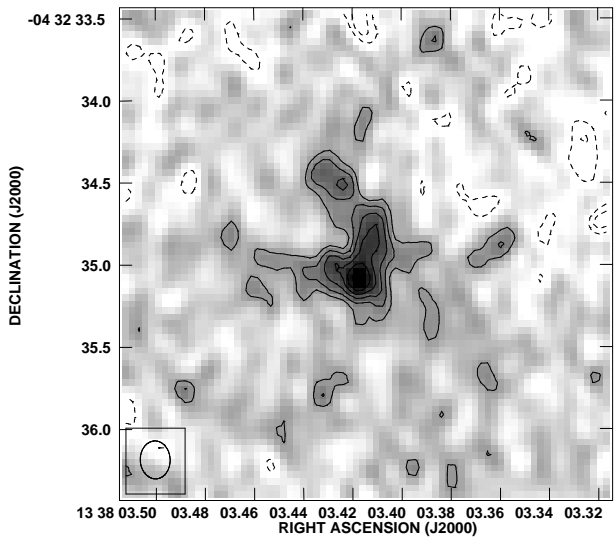}
\vspace{-10mm}

\caption{VLA map of the \bco\ emission toward BRI1335--0417 (integrated over the central 43.75\,MHz, or 308\,km\,s$^{-1}$), imaged using the combined B and C array dataset and natural weighting. 
Contours are shown at (-3, -2, 2, 3, 4, 5, 6, 7)$\times\sigma$
(1$\sigma = 50\,\mu$Jy beam$^{-1}$).
The beam size (0.23\,$''$$\times$0.18\,$''$)
is shown in the bottom left corner.
\label{f1}}
%
\end{figure}

We observed the \bco\ transition ($\nu_{\rm rest} = 230.53799\,$GHz)
towards \bri\ using the VLA in B configuration on 2005 April 11, 13,
and 22, and in C configuration on 2005 September 16. The total on-sky
integration time amounts to 29\,hr.  At $z=4.4074$, the line is
redshifted to 42.6338\,GHz (7.03\,mm).  Observations were performed in
fast-switching mode (e.g., Carilli \& Holdaway \citeyear{ch99}) using
the nearby source 13515-01513 for secondary amplitude and phase
calibration.  Observations were carried out under very good weather
conditions with 25 antennas. The phase stability in all
runs was excellent (typically $<$15$^\circ$ for the longest
baselines).
The phase coherence was checked by imaging the calibrator source
13569+02144 with the same calibration cycle as that used for the
target source.  For primary flux calibration, 3C\,286 was observed
during each run.  Given the restrictions of the VLA correlator, two
25\,MHz intermediate frequencies (IFs) with seven 3.125\,MHz channels
each were observed simultaneously centered at the \bco\ line
frequency, leading to an effective bandwidth of 43.75\,MHz
(corresponding to 308\kms\ at 42.6\,GHz). This encompasses a major
fraction of the CO line width as measured in the \eco\ transition
(420$\pm$60\,\kms\ FWHM, Guilloteau \etal\ \citeyear{gui97}), but does
not cover the line wings and the continuum. Earlier observations set a
2\,$\sigma$ limit of 240$\mu$Jy on the continuum emission (Carilli
\etal\ \citeyear{car99}), in agreement with an expected flux of 
$\sim$34\,$\mu$Jy as derived from the continuum spectral energy
distribution of this source.

\begin{figure}
\epsscale{1.15}
\vspace{-5mm}

\plotone{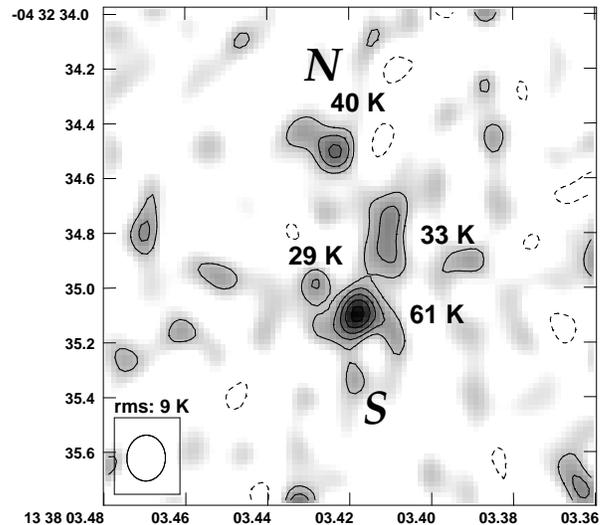}
\vspace{-10mm}

\caption{\bco\ emission line map of \bri\ at $\sim$1.0\,kpc resolution,
  imaged using B array data only and robust 0 weighting.
  Contours are shown at (--3, --2, 2, 3,
  4, 5, 6)$\times\sigma$ (1$\sigma = 58\,\mu$Jy beam$^{-1}$). The beam
  size (0.16\,$''$$\times$0.14\,$''$) is shown in the bottom left
  corner. The labels indicate the rest-frame peak brightness
  temperatures $T_{\rm b}$ in the brightest 
  molecular line emission regions.
\label{f2}}
%
\end{figure}

For data reduction and analysis, the AIPS
package was used. 
Two data sets were created for the final analysis. The first dataset
includes both the B- and C-array data, and is imaged using natural
weighting.  A velocity-integrated \bco\ map of this dataset is shown
in Fig.~\ref{f1}.  The synthesized clean beam has a size of
0.23\,$''$$\times$0.18\,$''$ (1.6$\times$1.2\,kpc).  The final rms
over the full bandwidth of 43.75\,MHz (308\,\kms) is 50\,$\mu$Jy
beam$^{-1}$.  To boost the resolution, the second dataset includes B
array data only, and are imaged using robust 0 (i.e., intermediate
between natural and uniform) weighting (Fig.~\ref{f2}), achieving a
resolution of 0.16\,$''$$\times$0.14\,$''$ (1.1$\times$0.95\,kpc).
This results in an rms of 58\,$\mu$Jy\,beam$^{-1}$ per channel.  In
Fig.~\ref{f3}, seven velocity channel maps (6.25\,MHz, or 44\,\kms\
each) of the \bco\ line based on the combined BC-array data are shown.
A Gaussian taper falling to 30\% level at a baseline length of
800\,k$\lambda$ (longest baseline in the B array data:
$\sim$1.6M$\lambda$) was applied to the UV data, resulting in a
resolution of 0.32\,$''$$\times$0.30\,$''$ (2.2$\times$2.0\,kpc), and
an rms of 150\,$\mu$Jy\,beam$^{-1}$ per channel. These channel maps
are combined into a three color overlay in Fig.~\ref{f4}.

\section{Results}

\subsection{Morphology of the Molecular Gas Reservoir}

\begin{figure*}
\epsscale{1.15}
\vspace{-10mm}

\plotone{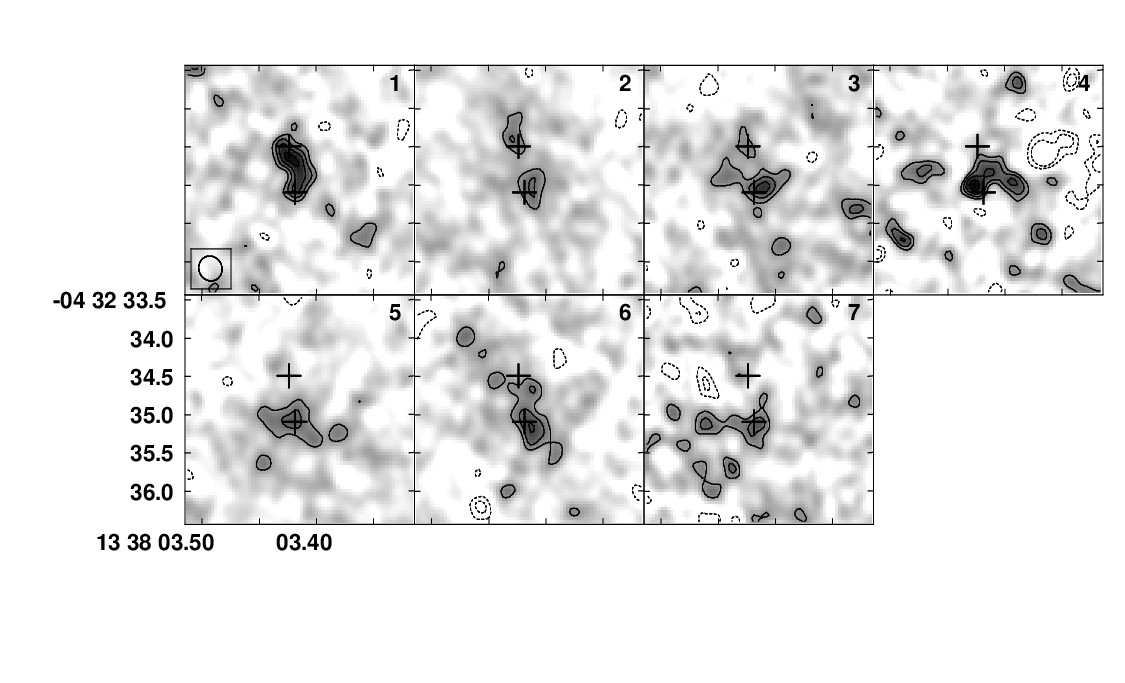}
\vspace{-20mm}

\caption{Channel maps of the \bco\ emission in \bri. The same region
  is shown as in Figure \ref{f1}. One channel width is 6.25\,MHz, or
  44\,\kms\ (frequencies increase with channel number and are shown at
  42615.050, 42621.300, 42627.550, 42633.800, 42640.050, 42646.300,
  and 42652.550\,MHz).  Contours are shown at (--3, --2, 2, 3, 4, 5)$\times \sigma$ (1$\sigma = 150\,\mu$Jy\,beam$^{-1}$). The
  beam size (tapered to 0.32\,$''$$\times$0.30\,$''$)
is shown in the bottom left corner; the crosses indicate 
main peaks of the the northern (`N') and southern (`S') CO components in
Fig.~\ref{f2}.
\label{f3}}
\end{figure*}

In Figure \ref{f1}, the integrated \bco\ emission over the full
measured bandpass (308\,\kms ) is shown at a linear resolution of
$\sim$1.4\,kpc (0.2$''$). The emission is clearly spatially resolved
over many beams, and extended out to a scale of 5\,kpc ($\sim$0.7$''$,
deconvolved for the beam size). Two distinct peaks of the emission are
identified, both of which are extended, and embedded in a continuous,
morphologically complex molecular structure. The northern component
has a peak flux density of 216 $\pm$ 50\,$\mu$Jy beam$^{-1}$, and the
southern component has a peak flux density of 388 $\pm$ 50\,$\mu$Jy
beam$^{-1}$ (peaks labeled `N' and `S' in Fig.~\ref{f2}). In Figure
\ref{f2}, a zoomed-in version of this map at 1.0\,kpc ($\sim$0.15$''$)
linear resolution is shown.
At this resolution, some of the more diffuse structure is resolved
out. Clearly, the southern component breaks up into multiple line
emission regions. Most notably, a resolved
subcomponent extends from the main southern peak (`S') toward the
northern peak (`N').
In the higher resolution map, the northern component has a peak flux
density of 246 $\pm$ 58\,$\mu$Jy beam$^{-1}$, and the southern
component has a peak flux density of 375 $\pm$ 58\,$\mu$Jy
beam$^{-1}$. The connecting component has a peak flux density of 202
$\pm$ 58\,$\mu$Jy beam$^{-1}$. These three peaks alone encompass a
molecular gas mass\footnote{As CO emission in currently known $z
\gtrsim 4$ quasar host galaxies appears to be thermalized typically 
up to $J$=4$\to$3 (Riechers et al.\ \citeyear{rie06}; Wei\ss\ et al.\
\citeyear{wei07}), constant $T_{\rm b}$ is assumed between \bco\ and
\aco.
The CO 
is more widespread than 
in nearby ultra-luminous infrared galaxies (ULIRGs), but shows similar
physical properties. We thus adopt a low ULIRG CO luminosity to H$_2$
mass conversion factor of
$\alpha$=0.8\,\msol\,(K\,\kms\,pc$^2$)$^{-1}$ (Downes \& Solomon
\citeyear{ds98}) rather than 
$\alpha$=4--5\,\msol\,(K\,\kms\,pc$^2$)$^{-1}$ as in nearby spiral
galaxies (e.g., Scoville \& Sanders \citeyear{sco87a}; Solomon
\& Barrett \citeyear{sol91}). Such low $\alpha$ are also found for $z
\sim 2.5$ submm galaxies (Tacconi et al.\ \citeyear{tac08}).}
of $M({\rm H_2})=4.1 \times 10^{10}\,$\msol, corresponding to about
two thirds of the mass of $M({\rm H_2})=6.6 \times 10^{10}\,$\msol\
derived from the integrated emission seen in the naturally weighted
dataset ($I_{\rm CO(2-1)}$=0.43$\pm$0.02\,Jy\,\kms, in agreement with
previous results by Carilli \etal\ \citeyear{car02}).
Accounting for the flux 
in the linewings that is not covered by our observations, the full
reservoir has an estimated mass of $M_{\rm tot}({\rm H_2})=9.2 \times
10^{10}\,$\msol. 

At $z$=4.4074, the brightest peaks of the northern and southern
components in the highest resolution map corresponds to beam-averaged,
rest-frame brightness temperatures of $T_{\rm b}$=61 $\pm$ 9\,K and 40
$\pm$ 9\,K (see Figure \ref{f2}). Averaging over the area of the full
\bco\ reservoir gives $T_{\rm b}$=10.0 $\pm$ 0.5\,K, which implies 
that part of the diffuse molecular structure may be rather cold.
Note that the minimum excitation temperature of \bco\ is $T_{\rm
ex}$=16.6\,K, which however is only 1.8\,K above the cosmic microwave
background (CMB) temperature at $z$=4.41.

\subsection{Dynamical Structure of the CO Distribution}

\begin{figure*}
\plotone{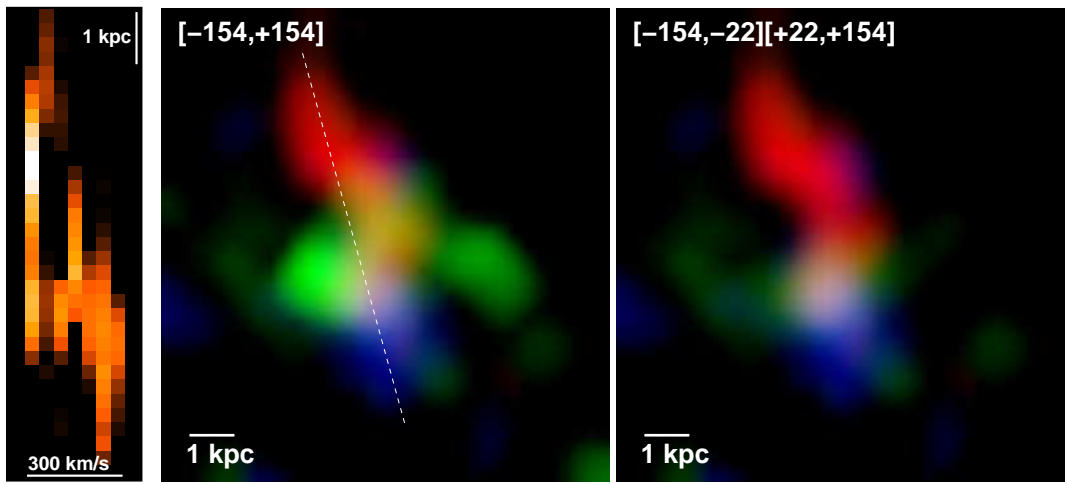}
\vspace{1.5mm}

\caption{RGB composite color map representation and position-velocity ($p-v$) diagram 
  of the \bco\ velocity channels shown in Fig.~\ref{f3}, with three
  colors encoding the velocity range of the emission [red: redshifted
  (channels 1-2 in Figure \ref{f3}), green: central (channels 3-5),
  blue: blueshifted (channels 6-7)]. The velocity range covered
  (relative to the line center) in \kms\ is indicated in the top left
  corner, and the linear scale is indicated in the bottom left corner
  of each panel. {\em Left}: $p-v$ diagram along a position axis of
  189$^\circ$, connecting the two brightest peaks of the integrated
  line emission.  {\em Middle}: All velocity channels (1-7). The
  dashed line indicates the orientation of the $p-v$ slice. {\em
  Right}: All except the central channel (channel 4).
  \label{f4}}
\end{figure*}

In Figure \ref{f3}, the \bco\ emission is shown in seven 44\,\kms\
wide velocity channels, smoothed to a linear resolution of
$\sim$2\,kpc. The emission is clearly dynamically resolved and is
moving, to first order, from north to south between the red and blue
velocity channels (i.e., with increasing channel number). In addition,
the center channel (\#4) shows a bright, compact peak, which
corresponds to a narrow velocity component. This peak is weighted down
in the integrated line maps, and corresponds to the third small
3$\sigma$ peak of the southern component in Figure \ref{f2} (labeled
with its $T_{\rm b}$ of 29 $\pm$ 9\,K). Figure
\ref{f4} shows a composite color map of the velocity channel maps and a position-velocity ($p-v$) diagram 
(left panel).  The middle panel includes all velocity channels, while
the central channel (covering the range from -22\,\kms\ to +22\,\kms)
is excluded from the right panel. The emission is clearly moving from
redshifted to blueshifted emission in the right panel. However, the
emission in the (green) central channels does not
follow the north-south extension, but also show an east-west
extension, indicating an additional, distinct velocity component in
the system. The overall peak of the emission where all components
overlap (seen as a `white spot' in the figure) corresponds to the
southern peak in the integrated line map.  The velocity structure thus
is more complex than that of a simple, inclined, rotating disk (as
also seen in the $p-v$ slice).
The central, compact, narrow velocity component 
is likely 
associated with the optical quasar.  However, the relative astrometry
of the radio and optical observations (Storrie-Lombardi \etal\
\citeyear{sl96}) is unfortunately not accurate enough at present to 
substantiate this conclusion.

\section{Discussion}

We present maps of the molecular gas distribution in a quasar host
galaxy at $z$=4.4 at 1.0\,kpc spatial resolution. The VLA data show
that the molecular gas reservoir in this galaxy is not only
distributed over a scale of $\sim$5\,kpc, but also structured in
velocity space.  This is the first time that the molecular gas in a
quasar host galaxy at such a high redshift (or, indeed, at any
redshift greater than $\sim$0) has ever been resolved both spatially
and dynamically over multiple beams (except for the lensed system
PSS\,J2322+1944 at $z$=4.12; Riechers et al.\ \citeyear{rie08}). The
emission in \bri\ is resolved into at least three distinct components
harboring at least 1--2 $\times$ 10$^{10}$\,\msol\ of molecular gas
each, and at least another 2 $\times$ 10$^{10}$\,\msol\ is found in
the more diffuse molecular medium in between these
concentrations. Each of the subcomponents hosts a few times the
molecular gas mass of nearby ULIRGs such as Arp\,220 (Downes \&
Solomon \citeyear{ds98}). The brightest peak corresponds to a 1\,kpc
diameter region with a gas surface density of $\Sigma_{\rm H_2}^{\rm
peak}$=2.3 $\times$ 10$^{10}$\,\msol\,kpc$^{-2}$, which is by a factor
of a few higher than in in the central regions of typical ULIRGs
(e.g., Wilson et al.\ \citeyear{wil08}).
The total molecular gas mass and distribution are reminiscent of those
in the $z$=4.69 and $z$=6.42 quasars BR\,1202--0725 and
SDSS\,J1148+5251 (Omont \etal\ \citeyear{omo96}; Carilli \etal\
\citeyear{car02}; Riechers \etal\ \citeyear{rie06}; 
Walter \etal\ \citeyear{wal04}), albeit concentrated
on almost an order of magnitude smaller scales than in BR\,1202--0725.
The brightest peak of the \bco\ emission traces a region
of 1\,kpc diameter where the gas has an average temperature of at
least 61\,K (depending on optical depth and beam dilution), which is
at the high end of but comparable to temperatures in the central
regions of nearby ULIRGs (e.g., Downes \& Solomon \citeyear{ds98}).
However, in ULIRGs, regions with molecular gas temperatures of
30--60\,K are much more compact than 1\,kpc.
The CO excitation ladder in \bri\ is not constrained well at
present. However, the excitation so far is consistent with being
similar to the $z$=4.69 quasar host galaxy of BR\,1202--0725, for
which models of collisional excitation predict a kinetic gas
temperature of $T_{\rm kin}$=60\,K and a median gas density of $n({\rm
H_2}) = 10^{4.1}\,$cm$^{-3}$ (Riechers et al.\
\citeyear{rie06}). If indeed $T_{\rm kin} \simeq T_{\rm b}^{\rm
peak}$, this would indicate that the brightest CO peak is close to
being resolved. Also, it appears that the gas temperature and density
are on the high end of but comparable to ULIRGs, while the total gas
mass is by about an order of magnitude higher.

Based on the FIR luminosity and SED shape of the source ($L_{\rm
FIR}$=3.1$\times$10$^{13}$\,\lsol; Benford et al.\ \citeyear{ben99}),
a star formation rate (SFR) of 4650\,\msol\,yr$^{-1}$ and a total dust
mass of 2$\times$10$^9$\,\msol\ can be derived.\footnote{Assuming
SFR=1.5$\times$10$^{-10}\,L_{\rm FIR}$(\msol\,yr$^{-1}$/\lsol)
(Kennicutt \citeyear{ken98a}).}
This does not account for possible heating of the
dust by the AGN. However, if the dust in this source has an extension
and non-uniform distribution similar to that of the molecular gas,
local heating (rather than heating of a central source) is likely to
be responsible for most of the dust emission. This is consistent with
the finding that the (rest-frame) 4.0\,cm radio continuum emission in
this source dominantly originates from extended, kpc-scale regions and
shows a peak $T_{\rm b}$ of only few times 10$^5$\,K (Momjian et al.\
\citeyear{mom07}). Such $T_{\rm b}$ are typical for relativistic
electrons from supernova remnants and the interstellar medium in
nuclear starbursts, but by at least 2 orders of magnitude lower than
found in AGN-powered environments. This finding is also reflected in
the fact that \bri\ follows the radio-FIR correlation for star-forming
galaxies (Carilli et al.\ \citeyear{car99}).  The FIR-derived SFR is
by about an order of magnitude higher than in local ULIRGs; however,
the ratio of SFR and total gas mass is consistent with the scaling
relation for ULIRGs (Solomon et al.\ \citeyear{sol97}) and other
high-$z$ galaxies (Solomon \& Vanden Bout \citeyear{sv05}; Riechers et
al.\ \citeyear{rie06}).
Moreover, in the centers of nearby ULIRGs, $\Sigma_{\rm H_2}^{\rm peak}$
correlates with the SFR (Wilson et al.\ \citeyear{wil08}), which
suggests that an increased availability of fuel for star formation at
a certain density leads to an increased SFR. Given its high
$\Sigma_{\rm H_2}^{\rm peak}$ on kpc scales, this also motivates the
high derived SFR for \bri. Assuming that the gas is converted into
stars at an efficiency of 5--10\% as in giant molecular cloud cores
(e.g., Myers et al.\
\citeyear{mye86}; Scoville et al.\ \citeyear{sco87}), the SFR
corresponds to a gas depletion timescale of 2--4 $\times$ 10$^8$\,yr.

If the molecular gas in this system was gravitationally bound, the CO
linewidth (see Sect.~2) and distribution would predict a dynamical
mass of $M_{\rm dyn}$=1.0 $\times$ 10$^{11}\,\sin^{-2}\,i$\,\msol,
which could account for both the total molecular gas mass and the mass
of the black hole of $M_{\rm BH}$=6 $\times$ 10$^9$\,\msol\ (Shields
\etal\
\citeyear{shi06}), but not for a substantial fraction of a
$\sim$4$\times$10$^{12}$\,\msol\ stellar bulge as predicted if the
local $M_{\rm BH}$--$\sigma_v$ relation were to hold (Ferrarese \&
Merritt \citeyear{fer00}; Gebhardt \etal\
\citeyear{geb00}).\footnote{In principle, assuming an extreme inclination 
toward the line-of-sight would significantly increase $M_{\rm dyn}$;
however, the implied large CO linewidths suggest that such a dynamical
structure would not be virialized (Riechers et al.\
\citeyear{rie08}).}

However, the overall structure of \bri\ looks rather disturbed, both
spatially and in its velocity structure. While the general north-south
extension of the source may be in agreement with 
rotating structure, this is not the case for the central part of the
emission, in particular the compact peak seen in the central channel.
This, and the structure seen in the high-resolution \bco\ map (Figure
\ref{f2}), are more reminiscent of the disturbed gas reservoirs in
major mergers, such as seen in the nearby Antennae (NGC\,4038/39,
e.g., Wilson \etal\ \citeyear{wil00}). If \bri\ was an interacting
system, the northern component may be in the process of merging with
the southern component, which likely hosts the luminous quasar. The
connecting component then may be the part where the two galaxies
overlap and merge. Note that this region alone would be massive enough
to host more than 10 of the largest molecular complexes found in the
overlap region of the Antennae (Wilson \etal\ \citeyear{wil00}).

Cosmological simulations predict that the merger rates at $z$=4.4 are
substantially higher than at $z$=0. Such scenarios imply that during
the early epoch of hierarchical galaxy formation, some of the most
massive galaxies form in major merger events (e.g., Springel \etal\
\citeyear{spr05}). Such major, `wet' mergers are believed to commonly
trigger both AGN and starburst activity, and lead to high excitation
of the molecular gas during both the hierarchical buildup of the host
galaxy and the quasar phase (e.g., Narayanan \etal\ \citeyear{nar07}).
In simulations, such objects often show multiple CO emisson peaks,
arising from molecular gas concentrations that have not yet fully
coalesced.

We conclude that the observed properties of \bri\ (AGN and extreme
starburst activity, high CO excitation, disturbed morphology over
5\,kpc scales) are connected to the ongoing buildup of the quasar host
galaxy.
Such a signpost of early galaxy assembly then could be considered
direct observational proof of the scenarios proposed by cosmological
simulations, and enable us to directly investigate the connection
between quasar activity and high-mass merger events at early cosmic
times.


\acknowledgments 
We thank the anonymous referee for valuable comments that helped to
improve the manuscript. DR acknowledges support from from NASA through
Hubble Fellowship grant HST-HF-01212.01-A awarded by the Space
Telescope Science Institute, which is operated by the Association of
Universities for Research in Astronomy, Inc., for NASA, under contract
NAS 5-26555, and from the Deutsche Forschungsgemeinschaft (DFG)
Priority Program 1177.  CC acknowledges support from the
Max-Planck-Gesellschaft and the Alexander von Humboldt-Stiftung
through the Max-Planck-Forschungspreis 2005.


\end{document}